\documentclass[twocolumn,prl,10pt,superscriptaddress]{revtex4-1}
\usepackage[latin1]{inputenc}
\usepackage[dvips]{graphicx}

\usepackage{dcolumn}
\usepackage{bm}
\usepackage{dsfont}
\usepackage{color}
\usepackage{url}
\usepackage[colorlinks=true ,citecolor=blue]{hyperref}
\usepackage{amsmath,amssymb,esint}
 \usepackage[table]{xcolor}

\definecolor{Nathanblue}{rgb}{0.,0.24,0.51}

\newcommand{\be}{\begin{equation}}
	\newcommand{\ee}{\end{equation}}
\newcommand{\bq}{\begin{eqnarray}}
	\newcommand{\eq}{\end{eqnarray}}

\newcommand\beq{\begin{equation}}
\newcommand\eeq{\end{equation}}
\newcommand\bal{\begin{aligned}}
\newcommand\eal{\end{aligned}}

\begin{document}

\title{Extended Nappi-Witten Geometry for the Fractional Quantum Hall Effect}

\author{Patricio Salgado-Rebolledo}
\affiliation{Universit\'e Libre de Bruxelles and International Solvay Institutes, ULB-Campus Plaine CP231, B-1050 Brussels, Belgium}

\author{Giandomenico Palumbo}
\affiliation{School of Theoretical Physics, Dublin Institute for Advanced Studies, 10 Burlington Road,
	Dublin 4, Ireland}

\date{\today}

\begin{abstract}

\noindent Motivated by the recent progresses in the formulation of geometric theories for the fractional quantum Hall states, we propose a novel non-relativistic geometric model for the Laughlin states based on an extension of the Nappi-Witten geometry. We show that 
the U(1) gauge sector responsible for the fractional Hall conductance, the gravitational Chern-Simons action and Wen-Zee term associated to the Hall viscosity can be derived from a single Chern-Simons theory with a gauge connection that takes values in the extended Nappi-Witten algebra. We then provide a new derivation of the chiral boson associated to the gapless edge states from the Wess-Zumino-Witten model that is induced by the Chern-Simons theory on the boundary.

\end{abstract}

\maketitle

\section{Introduction}

The Fractional Quantum Hall Effect (FQHE) is recognized as one of the most important physical phenomena in condensed matter physics~\cite{Laughlin,Haldane3,Jain,Fradkin4,Wen}. Being its microscopic origin a central research topic in topological phases of matter, it has given rise to an enormous amount of developments in quantum mesoscopic physics.
In the low energy regime, the fractional quantum Hall states (FQHs) can be described by Abelian and non-Abelian Chern-Simons theories, while the corresponding chiral edge states by rational conformal field theories (CFTs)~\cite{Wen,Wen2,Stone,Karabali,Cappelli2}. In the case of Laughlin states, the topological field theory depends on an emergent U(1) gauge field and the electromagnetic field, while the edge states are described by chiral bosons.
Recently, there have been an intense research on the geometric aspects of the FQHE. On one hand, 
the incompressibility of the FQHs is due to the presence of the strong external magnetic field and many-body interactions, which is encoded in the Girvin-Macdonald-Platzman (GMP) mode \cite{Girvin,Karabali,Cappelli4}. This mode can be naturally understood as a propagating non-relativistic spin-2 boson related to an emergent quantum geometry \cite{Haldane, Haldane2,Son5}. On the other hand, the background (ambient) geometry plays a central role in the Hall viscosity, which is a linear response effect of the Hall fluid in the bulk \cite{Avron,Read,Read-Rezayi,Hoyos-Son,Bradlyn4}. For all these reasons, several geometric models for the FQHE have been recently proposed \cite{Fradkin,Fradkin2,Bradlyn,Bradlyn2,Abanov-Gromov,Gromov-Abanov,Gromov-Jensen,Moroz,Cappelli,Geracie2,Wiegmann2,Wiegmann,Ferrari,Son6,Wu,Gromov-Son,Bradlyn3,Dwivedi}. These 2+1-dimensional effective field theories are based on non-relativistic geometry. In other words, the models are spatially covariant, and their formulation rely one the Newton-Cartan geometry \cite{Son6}.
This non-relativistic theory is a geometric reformulation of this Newton's gravity, that
imitates the geometric formulation of general relativity. It has been shown that Newton-Cartan theory
is based on the gauging of the Bargmann algebra (namely, centrally extended Galilei algebra) \cite{Andringa:2010it,Banerjee:2014pya,Bergshoeff:2017dqq}. 
Importantly, there have been many developments and generalizations of this theory by replacing the Bargmann algebra with the Newton-Hooke algebra, the Maxwell algebra, etc.
Behind these extensions of the Newton-Cartan theory, there appears the Nappi-Witten algebra \cite{Penafiel:2019czp}. This special algebra plays also an important role in certain Wess-Zumino-Witten (WZW) models and pp-wave spacetime \cite{Nappi-Witten}, and string-inspired 1+1-dimensional gravity~\cite{Jackiw2}.
Thus, generalized non-relativistic geometries can provide a novel scenario where to characterize the geometric features of FQHs. Notice, this research line follows in spirit the recent studies of the geometric aspects of topological insulators and topological superconductors where relativistic non-Riemannian geometries have been employed \cite{Palumbo,Palumbo2,Palumbo3,Palumbo4,Durka}.\\
In this work, we will present a novel geometric model for the Laughlin states given by a Chern-Simons (CS) theory with the gauge connection that takes values in an extended Nappi-Witten algebra. We will show that this CS action naturally contains not only the Wen-Zee \cite{Wen-Zee} and non-relativistic gravitational CS \cite{Fradkin} terms but also the U(1) topological terms responsible for the fractional Hall conductance \cite{Wen}. From this topological effective field theory we will derive the chiral WZW model on the boundary of the system. This CFT will allow us to describe the gapless edge states in terms of a chiral boson that takes contribution from both the charge and gravitational sectors. Finally, we will show that the extended Nappi-Witten geometry can be naturally embedded in a full non-relativistic space-time AdS-Lorentz geometry.
Our work paves the way for the characterization of FQHE from the the point of view of generalized non-relativistic geometries, where both the charge and gravitational sectors are encoded into a unified geometric formalism.


\section{Extended Nappi-Witten algebra}
The Nappi-Witten algebra is the central extension of the Euclidean algebra in two dimensions \cite{Nappi-Witten,Figueroa-OFarrill:1994liu}
\beq \label{NW}
\begin{array}{lll}
    \left[ P _a, P _b\right]=-\epsilon_{ab} T \,,\quad &
    \left[ J , P _a\right]=\epsilon_{a}^{\;\;b} P _b,
 \end{array}
\eeq
where $a= \{1,2\}$, $P_a$ stands for the translations in the two-dimensional plane, $J$ is the generator of rotations, and $T$ is central. Its Lorentzian version is isomorphic to the Newton-Hook algebra \cite{Alvarez:2007ys} as well as to the Maxwell algebra in two dimensions \cite{Afshar:2019axx}. The Maxwell algebra describes particle systems in the presence of a constant electromagnetic field \cite{Schrader:1972zd,Gomis:2017cmt}. Thus, the Nappi-Witten algebra has been found in the description of the integer QHE \cite{Duval:2000xr,Horvathy:2004am} because it naturally contains the magnetic translation algebra in two dimensions. Furthermore it has been shown that the Nappi-Witten geometry properly describes the momentum-space cigar geometry of a certain kind of two-dimensional topological phases \cite{Palumbo2}.

Based in the previous discussion, it seems natural to expect the Nappi-Witten algebra to be relevant in the description of two-dimensional interacting systems where a constant external electromagnetic field plays a central role. The quantum field theory of the Laughlin states is characterized by an extra field content given by an emergent gauge field $a$ \cite{Wen}. Thus, as a first attempt to describe these quantum states, we add an Abelian generator $Y$ to Eq.\eqref{NW} and consider the direct product Nappi-Witten$\times\mathfrak u(1)$. The central extension of the translations $T$ will be associated to the external electromagnetic field $A$ and the Abelian generator $Y$, to the emergent gauge field $a$. The Nappi-Witten$\times\mathfrak u(1)$ algebra admits a non-degenerate invariant bilinear form given by
\begin{equation}\label{invtensor}
\bal
\langle J  J \rangle&=\mu_0\,, \hskip.2truecm \langle P _a P _b\rangle=\mu_1\delta_{ab}\,,\hskip.2truecm \langle  J   T \rangle=-\mu_1\,,\\[6pt]
\langle Y  J \rangle&=  \rho_0\,, \hskip.2truecm
\langle Y Y \rangle=  \rho_1\,.
\eal
\end{equation}
The relevant gauge fields of the system are to encoded in a connection one-form $\mathbb A=\mathbb A_\mu^A T_A dx^\mu$ where $\mu=0,1,2$ is a space-time index and $T_A=\{J,P_a,Z,Y\}$ collectively denotes the generators of the Nappi-Witten$\times\mathfrak u(1)$ algebra. Explicitly, the gauge connection has the form
\beq\label{A1}
\mathbb A= \omega J+\frac{1}{\ell} e^a P_a   + A T + a Y.
\eeq
where $\omega$ is the connection of rotations (spin connection), $e^a$ the spatial dreibein and we have introduced a parameter $\ell$ with dimensions of length in such a way that the Lie algebra generators are dimensionless. We now introduce the following shift in the connection one-form in Eq.\eqref{A1}
\beq\label{shiftomega}
\omega \rightarrow \omega + \beta a,
\eeq
which can be translated into the definition of a new Lie algebra generator
\beq\label{newgen}
Z=Y+\beta J
\eeq
This leads to the following commutation relations
\beq \label{ExtNW}
\bal
	\left[ P _a, P _b\right]=-\epsilon_{ab} T \,,\quad &
	\left[ J , P _a\right]=\epsilon_{a}^{\;\;b} P _b 
	=\dfrac{1}{\beta} \left[ Z , P _a\right] \,,
\eal
\eeq
which we will refer to as \emph{extended Nappi-Witten algebra}. This algebra admits a non-degenerate invariant bilinear form given by
\begin{equation}\label{invtensor2}
\begin{array}{llll}
\langle J  J \rangle=\mu_0\,, \quad &\langle P _a P _b\rangle=\mu_1\delta_{ab}\,,\quad& \langle  J   T \rangle=-\mu_1\,.\\[6pt]
\langle Z  J \rangle= \mu_2\,, \quad &\langle ZT\rangle=-\beta \mu_1\,, \quad &\langle ZZ\rangle=\mu_3\, ,\\
\end{array}%
\end{equation}
where $\mu_i$ are real constants.
One can show that implementing the shift in Eq.\eqref{shiftomega} in the connection in Eq.\eqref{A1} and using the Nappi-Witten$\times\mathfrak u(1)$ algebra leads to the gauge connection for the extended Nappi-Witten algebra
\beq\label{A2}
\mathbb A= \omega J+\frac{1}{\ell} e^a P_a   + A T + a Z.
\eeq
The corresponding curvature $\mathbb F=(1/2)\mathbb F_{\mu\nu}^A T_A dx^\mu \wedge dx^\nu$ has the form
\beq
\mathbb F = d\omega J + \frac{1}{\ell}  R^a P_a + da Z + R T,
\eeq
where we have defined the one-forms
\beq
\bal
&R^a = de^a + \epsilon^a_{\;\;b}e^b( \omega +\beta a),\\
&R = dA +\frac{1}{2} \epsilon_{ab}e^a e^b.
\eal
\eeq
A gauge transformation $
\delta_\xi \mathbb A=d\xi +[ \mathbb A, \xi]
$ leads to the following transformation laws for the gauge fields
\beq
\bal
&
\delta_\xi \omega = d\xi^J\,,\quad
\delta_\xi a = d\xi^Z\,,\quad \delta_\xi A = d\xi^T+\frac{1}{\ell^2} \epsilon_{ab}\,\xi^{P_a}e^b 
\,,\\
&\delta_\xi e^a = d\xi^{P_a} + \epsilon^a_{\;\;b}
\Big(
\xi^J e^b -\omega \xi^{P_b}
+\beta\left( \xi^Z e^b - a \xi^{P_b}\right)
\Big)\,.
\eal
\eeq
Diffeomorphisms with parameter $\chi^\mu$ act on the connection as
\beq
\mathcal L_{\chi} \mathbb A_\mu= \mathbb F_{\mu\nu}\chi^\nu +\delta_{\xi_a \mathbb A^a}\mathbb A_\mu
\eeq
and thus the are on-shell equivalent to gauge transformations with parameter $\xi_a \mathbb A_\mu^a$.

\section{Topological Hall response from the Chern-Simons theory}

Because we are willing to derive the topological response of the fractional Hall states, in this section we consider the Chern-Simons action for the connection in Eq.\eqref{A2}, given by
\beq\label{CSaction}
S=-\frac{k}{4\pi}\int \left\langle\mathbb A\wedge d \mathbb A+ \dfrac{2}{3}\, \mathbb A\wedge\mathbb A\wedge\mathbb A \right\rangle,
\eeq
with $k$ an integer level. Importantly, this effective action can be derived from a microscopic theory by integrating out the fermionic fields associated to gapped spinful matter minimally coupled to the extended Nappi-Witten geometry (notice, spinful matter is compatible with a torsionful geometric background).
By employing Eq.\eqref{invtensor2}, the above action can be rewritten as follows (from now on the wedge product between forms will be omitted for simplicity)
\beq\label{CSactionENW}
\bal
S&=\displaystyle -\frac{k}{4\pi}\int \bigg[
\mu_0\,\omega d\omega   +
\frac{\mu_1}{\ell^2}  \,e^aR_a-2\mu_1\, Ad\omega\\
&\qquad\qquad
-2\mu_1\beta \,Ada+ 2\mu_2\, a d\omega +\mu_3 \, ada  \bigg]\,.
\eal
\eeq
Variation with respect to the spatial dreibein $e^a$ leads to the field equation $R^a=0$, which in turn yields the following equation for torsion
\beq \label{torsion}
T^a \equiv d e^a +\epsilon^a_{\;\;b} e^b \omega= -\beta \epsilon^a_{\;\;b} e^b a.
\eeq
This equation allows us to formally express the dreibein in terms of the spin connection and the field $a$.
Now,
by varying the action with respect to the field $a$, we obtain
\beq
a= \frac{\beta\mu_1}{\mu_3} A -\frac{\mu_2}{\mu_3} \omega.
\eeq
By replacing this expression in Eq.(\ref{CSactionENW}) and by taking the following identifications of the parameters
\begin{eqnarray}
k = \mu_3 =1,\hspace{0.3cm} \mu_1 = 2 \bar{s}\nu, \hspace{0.3cm} \mu_0 = \frac{c}{12}, \nonumber \\
 \mu_2 = \bar{s}\sqrt{\nu}, \hspace{0.3cm} \beta = \frac{1}{2 \bar{s}\sqrt{\nu}}, \hspace{0.8cm}
 \end{eqnarray}
we finally obtain
\beq
\bal
S=\displaystyle \int \bigg[
\left(\frac{\nu \bar{s}^2}{4 \pi}-\frac{c}{48 \pi}\right)\omega d\omega  
+\frac{\nu}{4 \pi}\,AdA
+\frac{\nu \bar{s}}{2 \pi}Ad\omega  \bigg],
\eal
\eeq
which is compatible with the effective geometric action for the FQHE analyzed in Refs.\cite{Fradkin,Fradkin2,Bradlyn,Bradlyn2,Abanov-Gromov,Gromov-Abanov,Gromov-Jensen,Moroz,Cappelli} although it is important to bear in mind that our spin connection is torsionful.
Here, $c$ is the chiral central charge, $\nu$ is the fractional filling and $\bar{s}$ is the average orbital spin. For the Laughlin states we have: $c=1$, $\nu= 1/(2p+1)$ and $\bar{s}=(2p+1)/2$, with $p$ an integer. The first and third terms in the action are usually referred as the gravitational Chern-Simons \cite{Fradkin} and Wen-Zee term \cite{Wen-Zee}, respectively.
Here, the coefficient in front of the U(1) CS term is associated to the Hall conductance, while the coefficient in front of the Wen-Zee term is related to the Hall viscosity \cite{Read,Read-Rezayi,Hoyos-Son,Bradlyn4,Golan-Hoyos}.
These are the two main topological responses in the Abelian FQH states on manifolds with genus 0 and 1 \cite{Wiegmann}. The electron Hall density and current derived from the above action are respectively given by
\begin{eqnarray}
\rho= \frac{\nu}{2 \pi}B + \frac{\nu \bar{s}}{4 \pi} R, \hspace{0.9cm} \nonumber\\
J^i= \frac{\nu}{2 \pi}\epsilon^{ij}E_j + \frac{\nu \bar{s}}{2 \pi}\epsilon^{ij} \mathcal{E}_j,
\end{eqnarray}
where $B$ and $E_j$ are the magnetic and electric fields, respectively while $R=(2/\sqrt{|e|}) \epsilon^{ij}\partial_i \omega_j$ and $\mathcal{E}_j=\partial_j\omega_0-\partial_0 \omega_j$ are the Abelian Ricci scalar and the gravi-electric field, respectively.
The Hall viscosity comes from the response of the system to shear or strain \cite{Avron}. Because our spin connection is torsionful, we follow here the approach developed in Refs.\cite{Geracie,Bradlyn} that allows us to define the symmetric Cauchy stress-mass tensor from the spin current $s^\mu$. This current is given by the variation of our topological action with respect to $\omega$
\begin{eqnarray}
s^\mu = \frac{\nu \bar{s}}{2 \pi}\epsilon^{\mu\nu\lambda}\partial_\nu A_\lambda+\left(\frac{\nu \bar{s}^2}{2 \pi}-\frac{c}{24 \pi}\right) \epsilon^{\mu\nu\lambda}\partial_\nu \omega_\lambda,
\end{eqnarray}
where the first term can be rewritten as follows
\begin{eqnarray}
\frac{\nu \bar{s}}{2 \pi}\epsilon^{\mu\nu\lambda}\partial_\nu A_\lambda = \eta_H u^{\mu},
\end{eqnarray}
with $u^{\mu} = (1, (1/B)\, \epsilon^{ij}E_j)$ the covariant drift velocity and $\eta_H=\nu \bar{s} B/4 \pi$  the Hall viscosity \cite{Geracie}.
Importantly, on a manifold with boundary both the Abelian and the gravitational CS terms contribute to the boundary gapless modes.
 These edge states are usually identified by chiral bosons. In the next section, we will derive the action for the chiral boson for the Laughlin states starting from the WZW action induced on the boundary by our CS theory in Eq.(\ref{CSaction}).

\section{Chiral boson from the WZW model}
It is well known that Chern-Simons theories are gauge invariant only on compact manifolds. For this reason, in this section we consider a manifold with boundary such that the CS action gives rise to a chiral WZW model on the boundary \cite{Dijkgraaf}. This CFT represents the natural effective theory for the gapless edge excitations of the FQH states. In the specific case of Laughlin states, the edge states are given by a chiral boson \cite{Wen,Wen2,Stone,Karabali,Cappelli2} that is usually described by the Floreanini-Jackiw action in flat space \cite{Jackiw}. In our framework, we now derive the chiral boson from a WZW model.
In fact, the CS action can be reduced to a boundary theory by solving the corresponding field equations for the spatial connection 
\begin{eqnarray}
\mathbb F_{ij}=0,
\end{eqnarray}
and then plugging the solution back in the action. A local solution is given by 
\beq
\mathbb A_i= g^{-1} \partial_i g.
\eeq
Imposing the condition
\beq
\mathbb A_t + v \mathbb A_\phi=0
\eeq
at the boundary and taking care of the surface integrals in \eqref{CSaction}, the action reduces to the following chiral WZW model \cite{Coussaert:1995zp}
\beq
\bal
S_{\rm WZW}&= 
\frac{k}{4\pi}\int_0^{2\pi} \hspace{-5pt} d\phi \int dt \, \left\langle g^{-1}\dot gg^{-1} g^\prime +v(g^{-1} g^\prime)^2 \right\rangle \\
&+\frac{k}{12\pi}\int_M\left\langle\left(g^{-1}{\rm d}g\right)^3\right\rangle,
\eal
\eeq
where $\dot g=\partial_t g$, $g^\prime=\partial_\phi g$ and we have chosen for simplicity a 2+1-dimensional manifold $M=D \times \mathbb R$, where $D$ is topologically equivalent to a disk and $\phi$ is the angular variable associated to the unitary circle that represents the spatial boundary $\mathbb S^1 \equiv \partial D$. Because we are mainly interested in the kinematics of the chiral boson, we look for the left-invariant Maurer-Cartan form and neglect the explicit derivation of the topological winding number of the WZW action.
One way to find the left-invariant Maurer-Cartan form \cite{Dijkgraaf} is to solve the corresponding Maurer-Cartan equation
\beq
d\Omega +\Omega\wedge\Omega=0,
\eeq
where
\beq
\Omega= g^{-1}dg= \Omega_J J  +\Omega_Z Z+ \Omega_P^a P_a +\Omega_T T.
\eeq
By employing the commutations of the extended Nappi-Witten algebra, the Maurer-Cartan equations can be shown to be equivalent to the following systems of equations
\beq
\bal
&d\Omega_J =0 \,,
\qquad d\Omega_Z =0\,,\\
&d\Omega_P^a+\epsilon^a_{\;\;b}  \Omega_P^b \left( \Omega_J + \beta \Omega_Z \right)
=0\,,\\
&d\Omega_T - \frac{1}{2}\epsilon_{ab}\Omega_P^a \Omega_P^b=0.
\eal
\eeq
The first two equations imply
\beq
\Omega_J= d \theta \,,\qquad
\Omega_Z= d \varphi,
\eeq
while for the others we have
\beq
\bal
\Omega_P^a&= d \sigma^a-\epsilon^a_{\;\;b}\sigma^b (d\theta+\beta d \varphi) \,,\\
\Omega_T&= d \vartheta+ \frac{1}{2}\epsilon_{ab}\sigma^a d\sigma^b +\frac{1}{2} \sigma^a \sigma_a (d\theta+\beta d \varphi),
\eal
\eeq
with $\theta$, $\varphi$ and $\vartheta$ three real scalar fields and $\sigma^a$ a vector field.
By defining the following coordinates
\beq
x^\pm = \frac{1}{2}\left(t+\frac{1}{v}\phi\right)\,,\qquad \partial_\pm= \partial_t +v\partial_\phi,
\eeq
with $v$ the Fermi velocity, we find the explicit form of the WZW action
\beq
\bal
&S_{WZW}= \int dt d \phi \bigg[ \mu_0 \partial_+  \theta \theta^\prime + 2\mu_2 \partial_+  \theta \varphi^\prime +\mu_3 \partial_+  \varphi \varphi^\prime \\
&
+\mu_1\Big(\partial_+ \sigma^a \sigma_a^\prime-  (2\partial_+\vartheta + \epsilon_{ab}\sigma^a \partial_+  \sigma^b) (\theta^\prime +\beta \varphi^\prime) \Big)
\bigg].
\eal
\eeq
We see that $\vartheta$ is a Lagrange multiplier which gives rise to the following constraint
\beq
\partial_+  \theta^\prime +\beta \partial_+  \varphi^\prime =0,
\eeq
which implies
\beq
\theta = -\beta \varphi + a(t) + b(x^-),
\eeq
with $a(t)$ and $b(x^-)$ arbitrary functions of their arguments.
On the other hand the field equation for $\sigma^a$ is given by
\beq
\partial_+ \sigma^\prime_a + \epsilon_{ab}\partial_+  \sigma^b (\theta^\prime+ \beta \varphi^\prime)=0.
\eeq
By replacing these expressions back in the action lead to an action for a single chiral boson, given by
\beq
S_{WZW}=\frac{c}{192\pi }\left(\frac{1}{\nu \bar{s}^2}\right) \int d\phi\, dt\ \left( \dot \varphi \varphi^{\prime}+v \varphi^\prime \varphi^\prime\right),
\eeq
which agrees with the boundary theory previously derived in literature \cite{Gromov-Jensen,Moroz} for the Laughlin states where $1/(\nu\bar{s}^2) = 4 \nu$ and $c=1$.
Importantly, the above chiral boson action can be naturally employed to describe the edge states of spin-$j$ Laughlin states \cite{Wiegmann,Wiegmann2,Ferrari} where
\beq
\bar{s}=\frac{1}{2\nu}-j.
\eeq

\section{AdS-Lorentz Algebra and Extended Newton-Cartan geometry}

The extended Nappi-Witten algebra can be naturally embedded in a full non-relativistic space-time algebra. This allows to interpret the Nappi-Witten geometry as a sub-manifold of a particular extended Newton-Cartan geometry \cite{Cartan:1923zea} . In 2+1 dimensions, the Galilei algebra admits two central extensions. This leads to the extended Bargmann symmetry \cite{levy1971galilei,Grigore:1993fz,Bose:1994sj,Bergshoeff:2016lwr}
\beq \label{Bargmann}
\begin{array}{lll}
    \left[ J , G _a\right]=\epsilon_{a}^{\;\;b} G _b\,,&
    \left[ J , P _a\right]= \left[ H , G _a\right]=\epsilon_{a}^{\;\;b} P _b\,,\\[5pt]
    \left[ G _a, G _b\right]=-\epsilon_{ab} S \,, &\left[ G _a, P _b\right]=-\epsilon_{ab} M \,.\\[5pt]
\end{array}
\eeq
One can consider the following an extension of the Bargmann algebra by including a new set of generators $\{Z,Z_a,T\}$ and the commutation relations
\beq \label{NRMaxwell}
\begin{array}{lll}
  &\left[ J , Z _a\right]=
    \left[ H , P _a\right]=
    \left[ Z , G _a\right]=\epsilon_a^{\;\;b} Z _b\,,\\[5pt]
   & \left[ P _a, P _b\right]=
    \left[ G _a, Z _b\right]=-\epsilon_{ab} T\,.\\[5pt]
\end{array}
\eeq
This is the Maxwellian Exotic  Bargmann algebra \cite{Aviles:2018jzw} and defines a central extension of the ``electric" non-relativistic Maxwell algebra \cite{Gonzalez:2016xwo,Gomis:2019fdh}. One can further extend the Bargmann algebra by introducing a parameter $\ell$ with dimension of length and the commutation relations \cite{Penafiel:2019czp}
\beq \label{NRAdSL}
\begin{array}{lll}
        \left[ Z , Z _a\right]= \epsilon_{a}^{\;\;b}  Z _b\,,&    \left[ H , Z _a\right]= \left[ Z , P _a\right]= \epsilon_{a}^{\;\;b}  P _b\,,     
\\[5pt]
\left[ P _a, Z _b\right]=- \epsilon_{ab}  M  \,, &
\left[ Z _a, Z _b\right]=- \epsilon_{ab} T  \,.
\end{array}
\eeq
This algebra can be obtained as a non-relativistic limit of the AdS-Lorentz algebra in 2+1 dimensions \cite{Concha:2019lhn}, which in turn is a semi-simple extension of the Maxwell algebra \cite{Soroka:2006aj}. As it happens in the relativistic case, the Maxwell and the AdS-Lorentz symmetries are related by an In\"on\"u-Wigner contraction. Indeed, one can use a length parameter $\ell$ to reinsert dimensions in the Lie algebra generators as
\beq
\bal
&H \rightarrow \ell H ,\quad P_a \rightarrow \ell P_a ,\quad\; M \rightarrow \ell M,
\\
&Z \rightarrow \ell^2 Z ,\quad Z_a \rightarrow \ell^2 Z_a
,\quad T \rightarrow \ell^2 T ,
\eal
\eeq
Thus, it is clear that in the limit $\ell\rightarrow\infty$ the non-relativistic AdS-Lorentz algebra reduces to the electric Maxwell symmetry. One can see that the extended Nappi-Witten algebra \eqref{ExtNW} is the sub-algebra of Eqs.\eqref{Bargmann}-\eqref{NRAdSL} spanned by the generators $\{J,P_a,Z,T\}$, where one has to use the redefinition $Z\rightarrow \beta Z$.

At the relativistic level, the Maxwell algebra has been previously used to construct a geometric model for the gapped boundary of three-dimensional topological insulators \cite{Palumbo}. However, in order to go beyond the integer QHE, a Chern-Simons term for the emergent U(1) gauge field is necessary. This term can be obtained by either extending the Maxwell to include its dual space, or by going to the AdS-Lorentz extension \cite{Durka}. Since the FQHE is intrinsically non-relativistic, it is natural to expect that the non-relativistic AdS-Lorentz is a good candidate to construct an effective geometric description. Here, we have shown that this is indeed the case.
A Chern-Simons action \eqref{CSaction} invariant under the non-relativistic AdS-Lorentz algebra is constructed by means of the connection one-form
\beq
\bal
\mathbb A&= \omega J+ \tau H+ a Z +\omega^a G_a + e^a P_a\\ &+ k^aZ_a + m M +s S+ A T .
\eal
\eeq
We can chose an absolute time and fix the reference frame by imposing the conditions
\beq
\tau_\mu = \delta_\mu^0\,,\qquad \omega^a_\mu =0
\eeq
Furthermore, we consider the particular case where
\beq
k^a_\mu=0\,, \qquad m_\mu=0=s_\mu \,.
\eeq
One can show that in this case the Chern-Simons action invariant under the non-relativistic AdS-Lorentz algebra reduces to the effective model in Eq.\eqref{CSactionENW}. It is important to note, however, that non-degenerate invariant bilinear form for the non-relativistic AdS-Lorentz symmetry that generalized \eqref{invtensor2} is given by
\begin{equation}\label{invtensorFULL}
\bal
&\langle  G _a  G _b\rangle=\lambda_0\delta_{ab}\,, \quad
\langle  G _a  P _b\rangle= \langle P _a Z _b\rangle =\lambda_1\delta_{ab} \\
& \langle P _a P _b\rangle= \langle G _a Z _b\rangle=\langle Z _a Z _b\rangle=\mu_1\delta_{ab}\\
& \langle  J   M  \rangle=\langle  H   S  \rangle=\langle H  T \rangle= \langle Z  M \rangle=-\lambda_1\,, \\
& \langle J  T \rangle=
  \langle Z  S \rangle=\langle H  M \rangle= \langle Z  T  \rangle=-\mu_1 \\
 &\langle  J   S \rangle=-\lambda_0\,,\; \langle  J   J \rangle=\mu_0\,,\; \langle  J   H \rangle=\lambda_4\,,\; \langle J  Z \rangle=\mu_2 \,,\\
&\langle H H \rangle=\lambda_2\,,\quad \langle  H   Z \rangle=\lambda_3\,,\quad   \langle Z  Z \rangle=\mu_3 \,.
\eal
\eeq
This expression is more general tan the invariant bilinear form derived in \cite{Penafiel:2019czp}, which is the one that comes from the relativistic AdS-Lorentz symmetry upon contraction. This indicates that the model here presented is purely non-relativistic and does not have a relativistic counterpart.

It is known that the gauging of the Bargmann algebra leads to Newton-Cartan geometry \cite{Andringa:2010it,Banerjee:2014pya,Bergshoeff:2017dqq}. Similarly, since the AdS-Lorentz algebra is an extension of the Bargmann symmetry, its gauging leads to an extended Newton-Cartan geometry. This is in complete analogy with the extended relativistic geometry that follows from Maxwell algebra \cite{Gibbons:2009me} (see also \cite{Salgado-Rebolledo:2019kft}), which is an extension of the Poincar\'e symmetry that underlies Minkowski space.

\section{Conclusions and outlook}
 In this paper, we have proposed a novel geometric model for the Laughlin states. We have shown that the U(1) and gravitational CS terms together with the Wen-Zee term can be derived from a single CS action where the gauge connection takes values in the extended Nappi-Witten algebra.
 Besides the topological response in the bulk given by the Hall conductance and Hall viscosity, we have provided a novel way to derive the effective field theory for the chiral boson that lives on the edge of the system. We have shown that the extended Nappi-Witten symmetry can be naturally embedded in the non-relativistic AdS-Lorentz algebra, which is a particular extension of the Bargmann algebra in 2+1 dimensions. In this way, the geometry behind our model can be thought as part of a generalized Newton-Cartan geometry. Several directions will be considered in future work. In particular, we will extend our formalism by including multi charged emergent gauge fields to describe FQH states beyond the Laughlin states \cite{Wen,Cappelli3} and a second emergent metric to properly encode the GMP mode and nematic states \cite{Gromov-Son,Bradlyn3}. Finally, we will define a novel non-relativistic higher-spin Nappi-Witten algebra to properly describe the higher-spin modes in the FQHE \cite{Son4,Papic}.  Our work paves the way for the description of the geometric and topological features of interacting topological fluids through generalized non-relativistic geometries where both the charge and gravitational sectors are dealt in a unified framework.

 {\bf Acknowledgments: }
The authors are pleased to acknowledge discussions with Barry Bradlyn.
This work was partially supported by FNRS-Belgium (conventions FRFC PDRT.1025.14 and  IISN 4.4503.15), as well as by funds from the Solvay Family.

\appendix

\bibliography{References}

\begin{thebibliography}{76}%
\makeatletter
\providecommand \@ifxundefined [1]{%
 \@ifx{#1\undefined}
}%
\providecommand \@ifnum [1]{%
 \ifnum #1\expandafter \@firstoftwo
 \else \expandafter \@secondoftwo
 \fi
}%
\providecommand \@ifx [1]{%
 \ifx #1\expandafter \@firstoftwo
 \else \expandafter \@secondoftwo
 \fi
}%
\providecommand \natexlab [1]{#1}%
\providecommand \enquote  [1]{``#1''}%
\providecommand \bibnamefont  [1]{#1}%
\providecommand \bibfnamefont [1]{#1}%
\providecommand \citenamefont [1]{#1}%
\providecommand \href@noop [0]{\@secondoftwo}%
\providecommand \href [0]{\begingroup \@sanitize@url \@href}%
\providecommand \@href[1]{\@@startlink{#1}\@@href}%
\providecommand \@@href[1]{\endgroup#1\@@endlink}%
\providecommand \@sanitize@url [0]{\catcode `\\12\catcode `\$12\catcode
  `\&12\catcode `\#12\catcode `\^12\catcode `\_12\catcode `\%12\relax}%
\providecommand \@@startlink[1]{}%
\providecommand \@@endlink[0]{}%
\providecommand \url  [0]{\begingroup\@sanitize@url \@url }%
\providecommand \@url [1]{\endgroup\@href {#1}{\urlprefix }}%
\providecommand \urlprefix  [0]{URL }%
\providecommand \Eprint [0]{\href }%
\providecommand \doibase [0]{http://dx.doi.org/}%
\providecommand \selectlanguage [0]{\@gobble}%
\providecommand \bibinfo  [0]{\@secondoftwo}%
\providecommand \bibfield  [0]{\@secondoftwo}%
\providecommand \translation [1]{[#1]}%
\providecommand \BibitemOpen [0]{}%
\providecommand \bibitemStop [0]{}%
\providecommand \bibitemNoStop [0]{.\EOS\space}%
\providecommand \EOS [0]{\spacefactor3000\relax}%
\providecommand \BibitemShut  [1]{\csname bibitem#1\endcsname}%
\let\auto@bib@innerbib\@empty
\bibitem [{\citenamefont {Laughlin}(1983)}]{Laughlin}%
  \BibitemOpen
  \bibfield  {author} {\bibinfo {author} {\bibfnamefont {R.~B.}\ \bibnamefont
  {Laughlin}},\ }\href {\doibase 10.1103/PhysRevLett.50.1395} {\bibfield
  {journal} {\bibinfo  {journal} {Phys. Rev. Lett.}\ }\textbf {\bibinfo
  {volume} {50}},\ \bibinfo {pages} {1395} (\bibinfo {year}
  {1983})}\BibitemShut {NoStop}%
\bibitem [{\citenamefont {Haldane}(1983)}]{Haldane3}%
  \BibitemOpen
  \bibfield  {author} {\bibinfo {author} {\bibfnamefont {F.~D.~M.}\
  \bibnamefont {Haldane}},\ }\href {\doibase 10.1103/PhysRevLett.51.605}
  {\bibfield  {journal} {\bibinfo  {journal} {Phys. Rev. Lett.}\ }\textbf
  {\bibinfo {volume} {51}},\ \bibinfo {pages} {605} (\bibinfo {year}
  {1983})}\BibitemShut {NoStop}%
\bibitem [{\citenamefont {Jain}(1989)}]{Jain}%
  \BibitemOpen
  \bibfield  {author} {\bibinfo {author} {\bibfnamefont {J.~K.}\ \bibnamefont
  {Jain}},\ }\href {\doibase 10.1103/PhysRevLett.63.199} {\bibfield  {journal}
  {\bibinfo  {journal} {Phys. Rev. Lett.}\ }\textbf {\bibinfo {volume} {63}},\
  \bibinfo {pages} {199} (\bibinfo {year} {1989})}\BibitemShut {NoStop}%
\bibitem [{\citenamefont {Lopez}\ and\ \citenamefont
  {Fradkin}(1991)}]{Fradkin4}%
  \BibitemOpen
  \bibfield  {author} {\bibinfo {author} {\bibfnamefont {A.}~\bibnamefont
  {Lopez}}\ and\ \bibinfo {author} {\bibfnamefont {E.}~\bibnamefont
  {Fradkin}},\ }\href {\doibase 10.1103/PhysRevB.44.5246} {\bibfield  {journal}
  {\bibinfo  {journal} {Phys. Rev. B}\ }\textbf {\bibinfo {volume} {44}},\
  \bibinfo {pages} {5246} (\bibinfo {year} {1991})}\BibitemShut {NoStop}%
\bibitem [{\citenamefont {Wen}(1992)}]{Wen}%
  \BibitemOpen
  \bibfield  {author} {\bibinfo {author} {\bibfnamefont {X.-G.}\ \bibnamefont
  {Wen}},\ }\href {\doibase 10.1142/S0217979292000840} {\bibfield  {journal}
  {\bibinfo  {journal} {International Journal of Modern Physics B}\ }\textbf
  {\bibinfo {volume} {06}},\ \bibinfo {pages} {1711} (\bibinfo {year}
  {1992})}\BibitemShut {NoStop}%
\bibitem [{\citenamefont {Wen}(1990)}]{Wen2}%
  \BibitemOpen
  \bibfield  {author} {\bibinfo {author} {\bibfnamefont {X.~G.}\ \bibnamefont
  {Wen}},\ }\href {\doibase 10.1103/PhysRevB.41.12838} {\bibfield  {journal}
  {\bibinfo  {journal} {Phys. Rev. B}\ }\textbf {\bibinfo {volume} {41}},\
  \bibinfo {pages} {12838} (\bibinfo {year} {1990})}\BibitemShut {NoStop}%
\bibitem [{\citenamefont {Stone}(1991)}]{Stone}%
  \BibitemOpen
  \bibfield  {author} {\bibinfo {author} {\bibfnamefont {M.}~\bibnamefont
  {Stone}},\ }\href {\doibase https://doi.org/10.1016/0003-4916(91)90177-A}
  {\bibfield  {journal} {\bibinfo  {journal} {Annals of Physics}\ }\textbf
  {\bibinfo {volume} {207}},\ \bibinfo {pages} {38 } (\bibinfo {year}
  {1991})}\BibitemShut {NoStop}%
\bibitem [{\citenamefont {Iso}\ \emph {et~al.}(1992)\citenamefont {Iso},
  \citenamefont {Karabali},\ and\ \citenamefont {Sakita}}]{Karabali}%
  \BibitemOpen
  \bibfield  {author} {\bibinfo {author} {\bibfnamefont {S.}~\bibnamefont
  {Iso}}, \bibinfo {author} {\bibfnamefont {D.}~\bibnamefont {Karabali}}, \
  and\ \bibinfo {author} {\bibfnamefont {B.}~\bibnamefont {Sakita}},\ }\href
  {\doibase https://doi.org/10.1016/0370-2693(92)90816-M} {\bibfield  {journal}
  {\bibinfo  {journal} {Physics Letters B}\ }\textbf {\bibinfo {volume}
  {296}},\ \bibinfo {pages} {143 } (\bibinfo {year} {1992})}\BibitemShut
  {NoStop}%
\bibitem [{\citenamefont {Cappelli}\ \emph
  {et~al.}(1993{\natexlab{a}})\citenamefont {Cappelli}, \citenamefont {Dunne},
  \citenamefont {Trugenberger},\ and\ \citenamefont {Zemba}}]{Cappelli2}%
  \BibitemOpen
  \bibfield  {author} {\bibinfo {author} {\bibfnamefont {A.}~\bibnamefont
  {Cappelli}}, \bibinfo {author} {\bibfnamefont {G.~V.}\ \bibnamefont {Dunne}},
  \bibinfo {author} {\bibfnamefont {C.~A.}\ \bibnamefont {Trugenberger}}, \
  and\ \bibinfo {author} {\bibfnamefont {G.~R.}\ \bibnamefont {Zemba}},\ }\href
  {\doibase https://doi.org/10.1016/0550-3213(93)90603-M} {\bibfield  {journal}
  {\bibinfo  {journal} {Nuclear Physics B}\ }\textbf {\bibinfo {volume}
  {398}},\ \bibinfo {pages} {531 } (\bibinfo {year}
  {1993}{\natexlab{a}})}\BibitemShut {NoStop}%
\bibitem [{\citenamefont {Girvin}\ \emph {et~al.}(1986)\citenamefont {Girvin},
  \citenamefont {MacDonald},\ and\ \citenamefont {Platzman}}]{Girvin}%
  \BibitemOpen
  \bibfield  {author} {\bibinfo {author} {\bibfnamefont {S.~M.}\ \bibnamefont
  {Girvin}}, \bibinfo {author} {\bibfnamefont {A.~H.}\ \bibnamefont
  {MacDonald}}, \ and\ \bibinfo {author} {\bibfnamefont {P.~M.}\ \bibnamefont
  {Platzman}},\ }\href {\doibase 10.1103/PhysRevB.33.2481} {\bibfield
  {journal} {\bibinfo  {journal} {Phys. Rev. B}\ }\textbf {\bibinfo {volume}
  {33}},\ \bibinfo {pages} {2481} (\bibinfo {year} {1986})}\BibitemShut
  {NoStop}%
\bibitem [{\citenamefont {Cappelli}\ \emph
  {et~al.}(1993{\natexlab{b}})\citenamefont {Cappelli}, \citenamefont
  {Trugenberger},\ and\ \citenamefont {Zemba}}]{Cappelli4}%
  \BibitemOpen
  \bibfield  {author} {\bibinfo {author} {\bibfnamefont {A.}~\bibnamefont
  {Cappelli}}, \bibinfo {author} {\bibfnamefont {C.~A.}\ \bibnamefont
  {Trugenberger}}, \ and\ \bibinfo {author} {\bibfnamefont {G.~R.}\
  \bibnamefont {Zemba}},\ }\href {\doibase
  https://doi.org/10.1016/0550-3213(93)90660-H} {\bibfield  {journal} {\bibinfo
   {journal} {Nuclear Physics B}\ }\textbf {\bibinfo {volume} {396}},\ \bibinfo
  {pages} {465 } (\bibinfo {year} {1993}{\natexlab{b}})}\BibitemShut {NoStop}%
\bibitem [{\citenamefont {Haldane}(2011)}]{Haldane}%
  \BibitemOpen
  \bibfield  {author} {\bibinfo {author} {\bibfnamefont {F.~D.~M.}\
  \bibnamefont {Haldane}},\ }\href {\doibase 10.1103/PhysRevLett.107.116801}
  {\bibfield  {journal} {\bibinfo  {journal} {Phys. Rev. Lett.}\ }\textbf
  {\bibinfo {volume} {107}},\ \bibinfo {pages} {116801} (\bibinfo {year}
  {2011})}\BibitemShut {NoStop}%
\bibitem [{\citenamefont {Yang}\ \emph {et~al.}(2012)\citenamefont {Yang},
  \citenamefont {Hu}, \citenamefont {Papi\ifmmode~\acute{c}\else \'{c}\fi{}},\
  and\ \citenamefont {Haldane}}]{Haldane2}%
  \BibitemOpen
  \bibfield  {author} {\bibinfo {author} {\bibfnamefont {B.}~\bibnamefont
  {Yang}}, \bibinfo {author} {\bibfnamefont {Z.-X.}\ \bibnamefont {Hu}},
  \bibinfo {author} {\bibfnamefont {Z.}~\bibnamefont
  {Papi\ifmmode~\acute{c}\else \'{c}\fi{}}}, \ and\ \bibinfo {author}
  {\bibfnamefont {F.~D.~M.}\ \bibnamefont {Haldane}},\ }\href {\doibase
  10.1103/PhysRevLett.108.256807} {\bibfield  {journal} {\bibinfo  {journal}
  {Phys. Rev. Lett.}\ }\textbf {\bibinfo {volume} {108}},\ \bibinfo {pages}
  {256807} (\bibinfo {year} {2012})}\BibitemShut {NoStop}%
\bibitem [{\citenamefont {Golkar}\ \emph
  {et~al.}(2016{\natexlab{a}})\citenamefont {Golkar}, \citenamefont {Nguyen},\
  and\ \citenamefont {Son}}]{Son5}%
  \BibitemOpen
  \bibfield  {author} {\bibinfo {author} {\bibfnamefont {S.}~\bibnamefont
  {Golkar}}, \bibinfo {author} {\bibfnamefont {D.~X.}\ \bibnamefont {Nguyen}},
  \ and\ \bibinfo {author} {\bibfnamefont {D.~T.}\ \bibnamefont {Son}},\ }\href
  {\doibase 10.1007/JHEP01(2016)021} {\bibfield  {journal} {\bibinfo  {journal}
  {Journal of High Energy Physics}\ }\textbf {\bibinfo {volume} {2016}},\
  \bibinfo {pages} {21} (\bibinfo {year} {2016}{\natexlab{a}})}\BibitemShut
  {NoStop}%
\bibitem [{\citenamefont {Avron}\ \emph {et~al.}(1995)\citenamefont {Avron},
  \citenamefont {Seiler},\ and\ \citenamefont {Zograf}}]{Avron}%
  \BibitemOpen
  \bibfield  {author} {\bibinfo {author} {\bibfnamefont {J.~E.}\ \bibnamefont
  {Avron}}, \bibinfo {author} {\bibfnamefont {R.}~\bibnamefont {Seiler}}, \
  and\ \bibinfo {author} {\bibfnamefont {P.~G.}\ \bibnamefont {Zograf}},\
  }\href {\doibase 10.1103/PhysRevLett.75.697} {\bibfield  {journal} {\bibinfo
  {journal} {Phys. Rev. Lett.}\ }\textbf {\bibinfo {volume} {75}},\ \bibinfo
  {pages} {697} (\bibinfo {year} {1995})}\BibitemShut {NoStop}%
\bibitem [{\citenamefont {Read}(2009)}]{Read}%
  \BibitemOpen
  \bibfield  {author} {\bibinfo {author} {\bibfnamefont {N.}~\bibnamefont
  {Read}},\ }\href {\doibase 10.1103/PhysRevB.79.045308} {\bibfield  {journal}
  {\bibinfo  {journal} {Phys. Rev. B}\ }\textbf {\bibinfo {volume} {79}},\
  \bibinfo {pages} {045308} (\bibinfo {year} {2009})}\BibitemShut {NoStop}%
\bibitem [{\citenamefont {Read}\ and\ \citenamefont
  {Rezayi}(2011)}]{Read-Rezayi}%
  \BibitemOpen
  \bibfield  {author} {\bibinfo {author} {\bibfnamefont {N.}~\bibnamefont
  {Read}}\ and\ \bibinfo {author} {\bibfnamefont {E.~H.}\ \bibnamefont
  {Rezayi}},\ }\href {\doibase 10.1103/PhysRevB.84.085316} {\bibfield
  {journal} {\bibinfo  {journal} {Phys. Rev. B}\ }\textbf {\bibinfo {volume}
  {84}},\ \bibinfo {pages} {085316} (\bibinfo {year} {2011})}\BibitemShut
  {NoStop}%
\bibitem [{\citenamefont {Hoyos}\ and\ \citenamefont {Son}(2012)}]{Hoyos-Son}%
  \BibitemOpen
  \bibfield  {author} {\bibinfo {author} {\bibfnamefont {C.}~\bibnamefont
  {Hoyos}}\ and\ \bibinfo {author} {\bibfnamefont {D.~T.}\ \bibnamefont
  {Son}},\ }\href {\doibase 10.1103/PhysRevLett.108.066805} {\bibfield
  {journal} {\bibinfo  {journal} {Phys. Rev. Lett.}\ }\textbf {\bibinfo
  {volume} {108}},\ \bibinfo {pages} {066805} (\bibinfo {year}
  {2012})}\BibitemShut {NoStop}%
\bibitem [{\citenamefont {Bradlyn}\ \emph {et~al.}(2012)\citenamefont
  {Bradlyn}, \citenamefont {Goldstein},\ and\ \citenamefont {Read}}]{Bradlyn4}%
  \BibitemOpen
  \bibfield  {author} {\bibinfo {author} {\bibfnamefont {B.}~\bibnamefont
  {Bradlyn}}, \bibinfo {author} {\bibfnamefont {M.}~\bibnamefont {Goldstein}},
  \ and\ \bibinfo {author} {\bibfnamefont {N.}~\bibnamefont {Read}},\ }\href
  {\doibase 10.1103/PhysRevB.86.245309} {\bibfield  {journal} {\bibinfo
  {journal} {Phys. Rev. B}\ }\textbf {\bibinfo {volume} {86}},\ \bibinfo
  {pages} {245309} (\bibinfo {year} {2012})}\BibitemShut {NoStop}%
\bibitem [{\citenamefont {Gromov}\ \emph {et~al.}(2015)\citenamefont {Gromov},
  \citenamefont {Cho}, \citenamefont {You}, \citenamefont {Abanov},\ and\
  \citenamefont {Fradkin}}]{Fradkin}%
  \BibitemOpen
  \bibfield  {author} {\bibinfo {author} {\bibfnamefont {A.}~\bibnamefont
  {Gromov}}, \bibinfo {author} {\bibfnamefont {G.~Y.}\ \bibnamefont {Cho}},
  \bibinfo {author} {\bibfnamefont {Y.}~\bibnamefont {You}}, \bibinfo {author}
  {\bibfnamefont {A.~G.}\ \bibnamefont {Abanov}}, \ and\ \bibinfo {author}
  {\bibfnamefont {E.}~\bibnamefont {Fradkin}},\ }\href {\doibase
  10.1103/PhysRevLett.114.016805} {\bibfield  {journal} {\bibinfo  {journal}
  {Phys. Rev. Lett.}\ }\textbf {\bibinfo {volume} {114}},\ \bibinfo {pages}
  {016805} (\bibinfo {year} {2015})}\BibitemShut {NoStop}%
\bibitem [{\citenamefont {Cho}\ \emph {et~al.}(2014)\citenamefont {Cho},
  \citenamefont {You},\ and\ \citenamefont {Fradkin}}]{Fradkin2}%
  \BibitemOpen
  \bibfield  {author} {\bibinfo {author} {\bibfnamefont {G.~Y.}\ \bibnamefont
  {Cho}}, \bibinfo {author} {\bibfnamefont {Y.}~\bibnamefont {You}}, \ and\
  \bibinfo {author} {\bibfnamefont {E.}~\bibnamefont {Fradkin}},\ }\href
  {\doibase 10.1103/PhysRevB.90.115139} {\bibfield  {journal} {\bibinfo
  {journal} {Phys. Rev. B}\ }\textbf {\bibinfo {volume} {90}},\ \bibinfo
  {pages} {115139} (\bibinfo {year} {2014})}\BibitemShut {NoStop}%
\bibitem [{\citenamefont {Bradlyn}\ and\ \citenamefont
  {Read}(2015{\natexlab{a}})}]{Bradlyn}%
  \BibitemOpen
  \bibfield  {author} {\bibinfo {author} {\bibfnamefont {B.}~\bibnamefont
  {Bradlyn}}\ and\ \bibinfo {author} {\bibfnamefont {N.}~\bibnamefont {Read}},\
  }\href {\doibase 10.1103/PhysRevB.91.125303} {\bibfield  {journal} {\bibinfo
  {journal} {Phys. Rev. B}\ }\textbf {\bibinfo {volume} {91}},\ \bibinfo
  {pages} {125303} (\bibinfo {year} {2015}{\natexlab{a}})}\BibitemShut
  {NoStop}%
\bibitem [{\citenamefont {Bradlyn}\ and\ \citenamefont
  {Read}(2015{\natexlab{b}})}]{Bradlyn2}%
  \BibitemOpen
  \bibfield  {author} {\bibinfo {author} {\bibfnamefont {B.}~\bibnamefont
  {Bradlyn}}\ and\ \bibinfo {author} {\bibfnamefont {N.}~\bibnamefont {Read}},\
  }\href {\doibase 10.1103/PhysRevB.91.165306} {\bibfield  {journal} {\bibinfo
  {journal} {Phys. Rev. B}\ }\textbf {\bibinfo {volume} {91}},\ \bibinfo
  {pages} {165306} (\bibinfo {year} {2015}{\natexlab{b}})}\BibitemShut
  {NoStop}%
\bibitem [{\citenamefont {Abanov}\ and\ \citenamefont
  {Gromov}(2014)}]{Abanov-Gromov}%
  \BibitemOpen
  \bibfield  {author} {\bibinfo {author} {\bibfnamefont {A.~G.}\ \bibnamefont
  {Abanov}}\ and\ \bibinfo {author} {\bibfnamefont {A.}~\bibnamefont
  {Gromov}},\ }\href {\doibase 10.1103/PhysRevB.90.014435} {\bibfield
  {journal} {\bibinfo  {journal} {Phys. Rev. B}\ }\textbf {\bibinfo {volume}
  {90}},\ \bibinfo {pages} {014435} (\bibinfo {year} {2014})}\BibitemShut
  {NoStop}%
\bibitem [{\citenamefont {Gromov}\ and\ \citenamefont
  {Abanov}(2014)}]{Gromov-Abanov}%
  \BibitemOpen
  \bibfield  {author} {\bibinfo {author} {\bibfnamefont {A.}~\bibnamefont
  {Gromov}}\ and\ \bibinfo {author} {\bibfnamefont {A.~G.}\ \bibnamefont
  {Abanov}},\ }\href {\doibase 10.1103/PhysRevLett.113.266802} {\bibfield
  {journal} {\bibinfo  {journal} {Phys. Rev. Lett.}\ }\textbf {\bibinfo
  {volume} {113}},\ \bibinfo {pages} {266802} (\bibinfo {year}
  {2014})}\BibitemShut {NoStop}%
\bibitem [{\citenamefont {Gromov}\ \emph {et~al.}(2016)\citenamefont {Gromov},
  \citenamefont {Jensen},\ and\ \citenamefont {Abanov}}]{Gromov-Jensen}%
  \BibitemOpen
  \bibfield  {author} {\bibinfo {author} {\bibfnamefont {A.}~\bibnamefont
  {Gromov}}, \bibinfo {author} {\bibfnamefont {K.}~\bibnamefont {Jensen}}, \
  and\ \bibinfo {author} {\bibfnamefont {A.~G.}\ \bibnamefont {Abanov}},\
  }\href {\doibase 10.1103/PhysRevLett.116.126802} {\bibfield  {journal}
  {\bibinfo  {journal} {Phys. Rev. Lett.}\ }\textbf {\bibinfo {volume} {116}},\
  \bibinfo {pages} {126802} (\bibinfo {year} {2016})}\BibitemShut {NoStop}%
\bibitem [{\citenamefont {Moroz}\ \emph {et~al.}(2015)\citenamefont {Moroz},
  \citenamefont {Hoyos},\ and\ \citenamefont {Radzihovsky}}]{Moroz}%
  \BibitemOpen
  \bibfield  {author} {\bibinfo {author} {\bibfnamefont {S.}~\bibnamefont
  {Moroz}}, \bibinfo {author} {\bibfnamefont {C.}~\bibnamefont {Hoyos}}, \ and\
  \bibinfo {author} {\bibfnamefont {L.}~\bibnamefont {Radzihovsky}},\ }\href
  {\doibase 10.1103/PhysRevB.91.195409} {\bibfield  {journal} {\bibinfo
  {journal} {Phys. Rev. B}\ }\textbf {\bibinfo {volume} {91}},\ \bibinfo
  {pages} {195409} (\bibinfo {year} {2015})}\BibitemShut {NoStop}%
\bibitem [{\citenamefont {Cappelli}\ and\ \citenamefont
  {Randellini}(2016)}]{Cappelli}%
  \BibitemOpen
  \bibfield  {author} {\bibinfo {author} {\bibfnamefont {A.}~\bibnamefont
  {Cappelli}}\ and\ \bibinfo {author} {\bibfnamefont {E.}~\bibnamefont
  {Randellini}},\ }\href {\doibase 10.1007/JHEP03(2016)105} {\bibfield
  {journal} {\bibinfo  {journal} {Journal of High Energy Physics}\ }\textbf
  {\bibinfo {volume} {2016}},\ \bibinfo {pages} {105} (\bibinfo {year}
  {2016})}\BibitemShut {NoStop}%
\bibitem [{\citenamefont {Geracie}\ \emph {et~al.}(2015)\citenamefont
  {Geracie}, \citenamefont {Son}, \citenamefont {Wu},\ and\ \citenamefont
  {Wu}}]{Geracie2}%
  \BibitemOpen
  \bibfield  {author} {\bibinfo {author} {\bibfnamefont {M.}~\bibnamefont
  {Geracie}}, \bibinfo {author} {\bibfnamefont {D.~T.}\ \bibnamefont {Son}},
  \bibinfo {author} {\bibfnamefont {C.}~\bibnamefont {Wu}}, \ and\ \bibinfo
  {author} {\bibfnamefont {S.-F.}\ \bibnamefont {Wu}},\ }\href {\doibase
  10.1103/PhysRevD.91.045030} {\bibfield  {journal} {\bibinfo  {journal} {Phys.
  Rev. D}\ }\textbf {\bibinfo {volume} {91}},\ \bibinfo {pages} {045030}
  (\bibinfo {year} {2015})}\BibitemShut {NoStop}%
\bibitem [{\citenamefont {Can}\ \emph {et~al.}(2014)\citenamefont {Can},
  \citenamefont {Laskin},\ and\ \citenamefont {Wiegmann}}]{Wiegmann2}%
  \BibitemOpen
  \bibfield  {author} {\bibinfo {author} {\bibfnamefont {T.}~\bibnamefont
  {Can}}, \bibinfo {author} {\bibfnamefont {M.}~\bibnamefont {Laskin}}, \ and\
  \bibinfo {author} {\bibfnamefont {P.}~\bibnamefont {Wiegmann}},\ }\href
  {\doibase 10.1103/PhysRevLett.113.046803} {\bibfield  {journal} {\bibinfo
  {journal} {Phys. Rev. Lett.}\ }\textbf {\bibinfo {volume} {113}},\ \bibinfo
  {pages} {046803} (\bibinfo {year} {2014})}\BibitemShut {NoStop}%
\bibitem [{\citenamefont {Klevtsov}\ and\ \citenamefont
  {Wiegmann}(2015)}]{Wiegmann}%
  \BibitemOpen
  \bibfield  {author} {\bibinfo {author} {\bibfnamefont {S.}~\bibnamefont
  {Klevtsov}}\ and\ \bibinfo {author} {\bibfnamefont {P.}~\bibnamefont
  {Wiegmann}},\ }\href {\doibase 10.1103/PhysRevLett.115.086801} {\bibfield
  {journal} {\bibinfo  {journal} {Phys. Rev. Lett.}\ }\textbf {\bibinfo
  {volume} {115}},\ \bibinfo {pages} {086801} (\bibinfo {year}
  {2015})}\BibitemShut {NoStop}%
\bibitem [{\citenamefont {Ferrari}\ and\ \citenamefont
  {Klevtsov}(2014)}]{Ferrari}%
  \BibitemOpen
  \bibfield  {author} {\bibinfo {author} {\bibfnamefont {F.}~\bibnamefont
  {Ferrari}}\ and\ \bibinfo {author} {\bibfnamefont {S.}~\bibnamefont
  {Klevtsov}},\ }\href {\doibase 10.1007/JHEP12(2014)086} {\bibfield  {journal}
  {\bibinfo  {journal} {Journal of High Energy Physics}\ }\textbf {\bibinfo
  {volume} {2014}},\ \bibinfo {pages} {86} (\bibinfo {year}
  {2014})}\BibitemShut {NoStop}%
\bibitem [{\citenamefont {Son}()}]{Son6}%
  \BibitemOpen
  \bibfield  {author} {\bibinfo {author} {\bibfnamefont {D.~T.}\ \bibnamefont
  {Son}},\ }\href {https://arxiv.org/abs/1306.0638} {\ }\Eprint
  {http://arxiv.org/abs/1306.0638} {arXiv:1306.0638} \BibitemShut {NoStop}%
\bibitem [{\citenamefont {Wu}\ and\ \citenamefont {Wu}(2015)}]{Wu}%
  \BibitemOpen
  \bibfield  {author} {\bibinfo {author} {\bibfnamefont {C.}~\bibnamefont
  {Wu}}\ and\ \bibinfo {author} {\bibfnamefont {S.-F.}\ \bibnamefont {Wu}},\
  }\href {\doibase 10.1007/JHEP01(2015)120} {\bibfield  {journal} {\bibinfo
  {journal} {Journal of High Energy Physics}\ }\textbf {\bibinfo {volume}
  {2015}},\ \bibinfo {pages} {120} (\bibinfo {year} {2015})}\BibitemShut
  {NoStop}%
\bibitem [{\citenamefont {Gromov}\ and\ \citenamefont
  {Son}(2017)}]{Gromov-Son}%
  \BibitemOpen
  \bibfield  {author} {\bibinfo {author} {\bibfnamefont {A.}~\bibnamefont
  {Gromov}}\ and\ \bibinfo {author} {\bibfnamefont {D.~T.}\ \bibnamefont
  {Son}},\ }\href {\doibase 10.1103/PhysRevX.7.041032} {\bibfield  {journal}
  {\bibinfo  {journal} {Phys. Rev. X}\ }\textbf {\bibinfo {volume} {7}},\
  \bibinfo {pages} {041032} (\bibinfo {year} {2017})}\BibitemShut {NoStop}%
\bibitem [{\citenamefont {Gromov}\ \emph {et~al.}(2017)\citenamefont {Gromov},
  \citenamefont {Geraedts},\ and\ \citenamefont {Bradlyn}}]{Bradlyn3}%
  \BibitemOpen
  \bibfield  {author} {\bibinfo {author} {\bibfnamefont {A.}~\bibnamefont
  {Gromov}}, \bibinfo {author} {\bibfnamefont {S.~D.}\ \bibnamefont
  {Geraedts}}, \ and\ \bibinfo {author} {\bibfnamefont {B.}~\bibnamefont
  {Bradlyn}},\ }\href {\doibase 10.1103/PhysRevLett.119.146602} {\bibfield
  {journal} {\bibinfo  {journal} {Phys. Rev. Lett.}\ }\textbf {\bibinfo
  {volume} {119}},\ \bibinfo {pages} {146602} (\bibinfo {year}
  {2017})}\BibitemShut {NoStop}%
\bibitem [{\citenamefont {Dwivedi}\ and\ \citenamefont
  {Klevtsov}(2019)}]{Dwivedi}%
  \BibitemOpen
  \bibfield  {author} {\bibinfo {author} {\bibfnamefont {V.}~\bibnamefont
  {Dwivedi}}\ and\ \bibinfo {author} {\bibfnamefont {S.}~\bibnamefont
  {Klevtsov}},\ }\href {\doibase 10.1103/PhysRevB.99.205158} {\bibfield
  {journal} {\bibinfo  {journal} {Phys. Rev. B}\ }\textbf {\bibinfo {volume}
  {99}},\ \bibinfo {pages} {205158} (\bibinfo {year} {2019})}\BibitemShut
  {NoStop}%
\bibitem [{\citenamefont {Andringa}\ \emph {et~al.}(2011)\citenamefont
  {Andringa}, \citenamefont {Bergshoeff}, \citenamefont {Panda},\ and\
  \citenamefont {de~Roo}}]{Andringa:2010it}%
  \BibitemOpen
  \bibfield  {author} {\bibinfo {author} {\bibfnamefont {R.}~\bibnamefont
  {Andringa}}, \bibinfo {author} {\bibfnamefont {E.}~\bibnamefont
  {Bergshoeff}}, \bibinfo {author} {\bibfnamefont {S.}~\bibnamefont {Panda}}, \
  and\ \bibinfo {author} {\bibfnamefont {M.}~\bibnamefont {de~Roo}},\ }\href
  {\doibase 10.1088/0264-9381/28/10/105011} {\bibfield  {journal} {\bibinfo
  {journal} {Class. Quant. Grav.}\ }\textbf {\bibinfo {volume} {28}},\ \bibinfo
  {pages} {105011} (\bibinfo {year} {2011})},\ \Eprint
  {http://arxiv.org/abs/1011.1145} {arXiv:1011.1145 [hep-th]} \BibitemShut
  {NoStop}%
\bibitem [{\citenamefont {Banerjee}\ \emph {et~al.}(2014)\citenamefont
  {Banerjee}, \citenamefont {Mitra},\ and\ \citenamefont
  {Mukherjee}}]{Banerjee:2014pya}%
  \BibitemOpen
  \bibfield  {author} {\bibinfo {author} {\bibfnamefont {R.}~\bibnamefont
  {Banerjee}}, \bibinfo {author} {\bibfnamefont {A.}~\bibnamefont {Mitra}}, \
  and\ \bibinfo {author} {\bibfnamefont {P.}~\bibnamefont {Mukherjee}},\ }\href
  {\doibase 10.1016/j.physletb.2014.09.004} {\bibfield  {journal} {\bibinfo
  {journal} {Phys. Lett. B}\ }\textbf {\bibinfo {volume} {737}},\ \bibinfo
  {pages} {369} (\bibinfo {year} {2014})},\ \Eprint
  {http://arxiv.org/abs/1404.4491} {arXiv:1404.4491 [gr-qc]} \BibitemShut
  {NoStop}%
\bibitem [{\citenamefont {Bergshoeff}\ \emph {et~al.}(2017)\citenamefont
  {Bergshoeff}, \citenamefont {Chatzistavrakidis}, \citenamefont {Romano},\
  and\ \citenamefont {Rosseel}}]{Bergshoeff:2017dqq}%
  \BibitemOpen
  \bibfield  {author} {\bibinfo {author} {\bibfnamefont {E.}~\bibnamefont
  {Bergshoeff}}, \bibinfo {author} {\bibfnamefont {A.}~\bibnamefont
  {Chatzistavrakidis}}, \bibinfo {author} {\bibfnamefont {L.}~\bibnamefont
  {Romano}}, \ and\ \bibinfo {author} {\bibfnamefont {J.}~\bibnamefont
  {Rosseel}},\ }\href {\doibase 10.1007/JHEP10(2017)194} {\bibfield  {journal}
  {\bibinfo  {journal} {JHEP}\ }\textbf {\bibinfo {volume} {10}},\ \bibinfo
  {pages} {194} (\bibinfo {year} {2017})},\ \Eprint
  {http://arxiv.org/abs/1708.05414} {arXiv:1708.05414 [hep-th]} \BibitemShut
  {NoStop}%
\bibitem [{\citenamefont {Pe\~nafiel}\ and\ \citenamefont
  {Salgado-Rebolledo}(2019)}]{Penafiel:2019czp}%
  \BibitemOpen
  \bibfield  {author} {\bibinfo {author} {\bibfnamefont {D.~M.}\ \bibnamefont
  {Pe\~nafiel}}\ and\ \bibinfo {author} {\bibfnamefont {P.}~\bibnamefont
  {Salgado-Rebolledo}},\ }\href {\doibase 10.1016/j.physletb.2019.135005}
  {\bibfield  {journal} {\bibinfo  {journal} {Phys. Lett. B}\ }\textbf
  {\bibinfo {volume} {798}},\ \bibinfo {pages} {135005} (\bibinfo {year}
  {2019})},\ \Eprint {http://arxiv.org/abs/1906.02161} {arXiv:1906.02161
  [hep-th]} \BibitemShut {NoStop}%
\bibitem [{\citenamefont {Nappi}\ and\ \citenamefont
  {Witten}(1993)}]{Nappi-Witten}%
  \BibitemOpen
  \bibfield  {author} {\bibinfo {author} {\bibfnamefont {C.~R.}\ \bibnamefont
  {Nappi}}\ and\ \bibinfo {author} {\bibfnamefont {E.}~\bibnamefont {Witten}},\
  }\href {\doibase 10.1103/PhysRevLett.71.3751} {\bibfield  {journal} {\bibinfo
   {journal} {Phys. Rev. Lett.}\ }\textbf {\bibinfo {volume} {71}},\ \bibinfo
  {pages} {3751} (\bibinfo {year} {1993})}\BibitemShut {NoStop}%
\bibitem [{\citenamefont {Cangemi}\ and\ \citenamefont
  {Jackiw}(1992)}]{Jackiw2}%
  \BibitemOpen
  \bibfield  {author} {\bibinfo {author} {\bibfnamefont {D.}~\bibnamefont
  {Cangemi}}\ and\ \bibinfo {author} {\bibfnamefont {R.}~\bibnamefont
  {Jackiw}},\ }\href {\doibase 10.1103/PhysRevLett.69.233} {\bibfield
  {journal} {\bibinfo  {journal} {Phys. Rev. Lett.}\ }\textbf {\bibinfo
  {volume} {69}},\ \bibinfo {pages} {233} (\bibinfo {year} {1992})}\BibitemShut
  {NoStop}%
\bibitem [{\citenamefont {Palumbo}(2017)}]{Palumbo}%
  \BibitemOpen
  \bibfield  {author} {\bibinfo {author} {\bibfnamefont {G.}~\bibnamefont
  {Palumbo}},\ }\href {\doibase https://doi.org/10.1016/j.aop.2017.08.018}
  {\bibfield  {journal} {\bibinfo  {journal} {Annals of Physics}\ }\textbf
  {\bibinfo {volume} {386}},\ \bibinfo {pages} {15 } (\bibinfo {year}
  {2017})}\BibitemShut {NoStop}%
\bibitem [{\citenamefont {Palumbo}(2018)}]{Palumbo2}%
  \BibitemOpen
  \bibfield  {author} {\bibinfo {author} {\bibfnamefont {G.}~\bibnamefont
  {Palumbo}},\ }\href {\doibase 10.1140/epjp/i2018-11856-8} {\bibfield
  {journal} {\bibinfo  {journal} {The European Physical Journal Plus}\ }\textbf
  {\bibinfo {volume} {133}},\ \bibinfo {pages} {23} (\bibinfo {year}
  {2018})}\BibitemShut {NoStop}%
\bibitem [{\citenamefont {Palumbo}\ and\ \citenamefont
  {Pachos}(2016)}]{Palumbo3}%
  \BibitemOpen
  \bibfield  {author} {\bibinfo {author} {\bibfnamefont {G.}~\bibnamefont
  {Palumbo}}\ and\ \bibinfo {author} {\bibfnamefont {J.~K.}\ \bibnamefont
  {Pachos}},\ }\href {\doibase https://doi.org/10.1016/j.aop.2016.05.005}
  {\bibfield  {journal} {\bibinfo  {journal} {Annals of Physics}\ }\textbf
  {\bibinfo {volume} {372}},\ \bibinfo {pages} {175 } (\bibinfo {year}
  {2016})}\BibitemShut {NoStop}%
\bibitem [{\citenamefont {Maraner}\ \emph {et~al.}(2019)\citenamefont
  {Maraner}, \citenamefont {Pachos},\ and\ \citenamefont {Palumbo}}]{Palumbo4}%
  \BibitemOpen
  \bibfield  {author} {\bibinfo {author} {\bibfnamefont {P.}~\bibnamefont
  {Maraner}}, \bibinfo {author} {\bibfnamefont {J.~K.}\ \bibnamefont {Pachos}},
  \ and\ \bibinfo {author} {\bibfnamefont {G.}~\bibnamefont {Palumbo}},\ }\href
  {\doibase 10.1038/s41598-019-53771-5} {\bibfield  {journal} {\bibinfo
  {journal} {Scientific Reports}\ }\textbf {\bibinfo {volume} {9}},\ \bibinfo
  {pages} {17308} (\bibinfo {year} {2019})}\BibitemShut {NoStop}%
\bibitem [{\citenamefont {Durka}\ and\ \citenamefont
  {Kowalski-Glikman}(2019)}]{Durka}%
  \BibitemOpen
  \bibfield  {author} {\bibinfo {author} {\bibfnamefont {R.}~\bibnamefont
  {Durka}}\ and\ \bibinfo {author} {\bibfnamefont {J.}~\bibnamefont
  {Kowalski-Glikman}},\ }\href {\doibase
  https://doi.org/10.1016/j.physletb.2019.06.058} {\bibfield  {journal}
  {\bibinfo  {journal} {Physics Letters B}\ }\textbf {\bibinfo {volume}
  {795}},\ \bibinfo {pages} {516 } (\bibinfo {year} {2019})}\BibitemShut
  {NoStop}%
\bibitem [{\citenamefont {Wen}\ and\ \citenamefont {Zee}(1992)}]{Wen-Zee}%
  \BibitemOpen
  \bibfield  {author} {\bibinfo {author} {\bibfnamefont {X.~G.}\ \bibnamefont
  {Wen}}\ and\ \bibinfo {author} {\bibfnamefont {A.}~\bibnamefont {Zee}},\
  }\href {\doibase 10.1103/PhysRevLett.69.953} {\bibfield  {journal} {\bibinfo
  {journal} {Phys. Rev. Lett.}\ }\textbf {\bibinfo {volume} {69}},\ \bibinfo
  {pages} {953} (\bibinfo {year} {1992})}\BibitemShut {NoStop}%
\bibitem [{\citenamefont {Figueroa-O'Farrill}\ and\ \citenamefont
  {Stanciu}(1994)}]{Figueroa-OFarrill:1994liu}%
  \BibitemOpen
  \bibfield  {author} {\bibinfo {author} {\bibfnamefont {J.~M.}\ \bibnamefont
  {Figueroa-O'Farrill}}\ and\ \bibinfo {author} {\bibfnamefont
  {S.}~\bibnamefont {Stanciu}},\ }\href {\doibase 10.1016/0370-2693(94)91525-3}
  {\bibfield  {journal} {\bibinfo  {journal} {Phys. Lett. B}\ }\textbf
  {\bibinfo {volume} {327}},\ \bibinfo {pages} {40} (\bibinfo {year} {1994})},\
  \Eprint {http://arxiv.org/abs/hep-th/9402035} {arXiv:hep-th/9402035}
  \BibitemShut {NoStop}%
\bibitem [{\citenamefont {Alvarez}\ \emph {et~al.}(2008)\citenamefont
  {Alvarez}, \citenamefont {Gomis}, \citenamefont {Kamimura},\ and\
  \citenamefont {Plyushchay}}]{Alvarez:2007ys}%
  \BibitemOpen
  \bibfield  {author} {\bibinfo {author} {\bibfnamefont {P.~D.}\ \bibnamefont
  {Alvarez}}, \bibinfo {author} {\bibfnamefont {J.}~\bibnamefont {Gomis}},
  \bibinfo {author} {\bibfnamefont {K.}~\bibnamefont {Kamimura}}, \ and\
  \bibinfo {author} {\bibfnamefont {M.~S.}\ \bibnamefont {Plyushchay}},\ }\href
  {\doibase 10.1016/j.physletb.2007.12.016} {\bibfield  {journal} {\bibinfo
  {journal} {Phys. Lett. B}\ }\textbf {\bibinfo {volume} {659}},\ \bibinfo
  {pages} {906} (\bibinfo {year} {2008})},\ \Eprint
  {http://arxiv.org/abs/0711.2644} {arXiv:0711.2644 [hep-th]} \BibitemShut
  {NoStop}%
\bibitem [{\citenamefont {Afshar}\ \emph {et~al.}(2020)\citenamefont {Afshar},
  \citenamefont {Gonz\'alez}, \citenamefont {Grumiller},\ and\ \citenamefont
  {Vassilevich}}]{Afshar:2019axx}%
  \BibitemOpen
  \bibfield  {author} {\bibinfo {author} {\bibfnamefont {H.}~\bibnamefont
  {Afshar}}, \bibinfo {author} {\bibfnamefont {H.~A.}\ \bibnamefont
  {Gonz\'alez}}, \bibinfo {author} {\bibfnamefont {D.}~\bibnamefont
  {Grumiller}}, \ and\ \bibinfo {author} {\bibfnamefont {D.}~\bibnamefont
  {Vassilevich}},\ }\href {\doibase 10.1103/PhysRevD.101.086024} {\bibfield
  {journal} {\bibinfo  {journal} {Phys. Rev. D}\ }\textbf {\bibinfo {volume}
  {101}},\ \bibinfo {pages} {086024} (\bibinfo {year} {2020})},\ \Eprint
  {http://arxiv.org/abs/1911.05739} {arXiv:1911.05739 [hep-th]} \BibitemShut
  {NoStop}%
\bibitem [{\citenamefont {Schrader}(1972)}]{Schrader:1972zd}%
  \BibitemOpen
  \bibfield  {author} {\bibinfo {author} {\bibfnamefont {R.}~\bibnamefont
  {Schrader}},\ }\href {\doibase 10.1002/prop.19720201202} {\bibfield
  {journal} {\bibinfo  {journal} {Fortsch. Phys.}\ }\textbf {\bibinfo {volume}
  {20}},\ \bibinfo {pages} {701} (\bibinfo {year} {1972})}\BibitemShut
  {NoStop}%
\bibitem [{\citenamefont {Gomis}\ and\ \citenamefont
  {Kleinschmidt}(2017)}]{Gomis:2017cmt}%
  \BibitemOpen
  \bibfield  {author} {\bibinfo {author} {\bibfnamefont {J.}~\bibnamefont
  {Gomis}}\ and\ \bibinfo {author} {\bibfnamefont {A.}~\bibnamefont
  {Kleinschmidt}},\ }\href {\doibase 10.1007/JHEP07(2017)085} {\bibfield
  {journal} {\bibinfo  {journal} {JHEP}\ }\textbf {\bibinfo {volume} {07}},\
  \bibinfo {pages} {085} (\bibinfo {year} {2017})},\ \Eprint
  {http://arxiv.org/abs/1705.05854} {arXiv:1705.05854 [hep-th]} \BibitemShut
  {NoStop}%
\bibitem [{\citenamefont {Duval}\ and\ \citenamefont
  {Horvathy}(2000)}]{Duval:2000xr}%
  \BibitemOpen
  \bibfield  {author} {\bibinfo {author} {\bibfnamefont {C.}~\bibnamefont
  {Duval}}\ and\ \bibinfo {author} {\bibfnamefont {P.~A.}\ \bibnamefont
  {Horvathy}},\ }\href {\doibase 10.1016/S0370-2693(00)00341-5} {\bibfield
  {journal} {\bibinfo  {journal} {Phys. Lett. B}\ }\textbf {\bibinfo {volume}
  {479}},\ \bibinfo {pages} {284} (\bibinfo {year} {2000})},\ \Eprint
  {http://arxiv.org/abs/hep-th/0002233} {arXiv:hep-th/0002233} \BibitemShut
  {NoStop}%
\bibitem [{\citenamefont {Horvathy}\ \emph {et~al.}(2005)\citenamefont
  {Horvathy}, \citenamefont {Martina},\ and\ \citenamefont
  {Stichel}}]{Horvathy:2004am}%
  \BibitemOpen
  \bibfield  {author} {\bibinfo {author} {\bibfnamefont {P.~A.}\ \bibnamefont
  {Horvathy}}, \bibinfo {author} {\bibfnamefont {L.}~\bibnamefont {Martina}}, \
  and\ \bibinfo {author} {\bibfnamefont {P.~C.}\ \bibnamefont {Stichel}},\
  }\href {\doibase 10.1016/j.physletb.2005.04.004} {\bibfield  {journal}
  {\bibinfo  {journal} {Phys. Lett. B}\ }\textbf {\bibinfo {volume} {615}},\
  \bibinfo {pages} {87} (\bibinfo {year} {2005})},\ \Eprint
  {http://arxiv.org/abs/hep-th/0412090} {arXiv:hep-th/0412090} \BibitemShut
  {NoStop}%
\bibitem [{\citenamefont {Golan}\ \emph {et~al.}(2019)\citenamefont {Golan},
  \citenamefont {Hoyos},\ and\ \citenamefont {Moroz}}]{Golan-Hoyos}%
  \BibitemOpen
  \bibfield  {author} {\bibinfo {author} {\bibfnamefont {O.}~\bibnamefont
  {Golan}}, \bibinfo {author} {\bibfnamefont {C.}~\bibnamefont {Hoyos}}, \ and\
  \bibinfo {author} {\bibfnamefont {S.}~\bibnamefont {Moroz}},\ }\href
  {\doibase 10.1103/PhysRevB.100.104512} {\bibfield  {journal} {\bibinfo
  {journal} {Phys. Rev. B}\ }\textbf {\bibinfo {volume} {100}},\ \bibinfo
  {pages} {104512} (\bibinfo {year} {2019})}\BibitemShut {NoStop}%
\bibitem [{\citenamefont {Geracie}\ \emph {et~al.}(2017)\citenamefont
  {Geracie}, \citenamefont {Prabhu},\ and\ \citenamefont {Roberts}}]{Geracie}%
  \BibitemOpen
  \bibfield  {author} {\bibinfo {author} {\bibfnamefont {M.}~\bibnamefont
  {Geracie}}, \bibinfo {author} {\bibfnamefont {K.}~\bibnamefont {Prabhu}}, \
  and\ \bibinfo {author} {\bibfnamefont {M.~M.}\ \bibnamefont {Roberts}},\
  }\href {\doibase https://doi.org/10.1007/JHEP06(2017)089} {\bibfield
  {journal} {\bibinfo  {journal} {Journal of High Energy Physics}\ }\textbf
  {\bibinfo {volume} {2017}},\ \bibinfo {pages} {89} (\bibinfo {year}
  {2017})}\BibitemShut {NoStop}%
\bibitem [{\citenamefont {Dijkgraaf}\ and\ \citenamefont
  {Witten}(1990)}]{Dijkgraaf}%
  \BibitemOpen
  \bibfield  {author} {\bibinfo {author} {\bibfnamefont {R.}~\bibnamefont
  {Dijkgraaf}}\ and\ \bibinfo {author} {\bibfnamefont {E.}~\bibnamefont
  {Witten}},\ }\href {\doibase https://doi.org/10.1007/BF02096988} {\bibfield
  {journal} {\bibinfo  {journal} {Communications in Mathematical Physics}\
  }\textbf {\bibinfo {volume} {129}},\ \bibinfo {pages} {393} (\bibinfo {year}
  {1990})}\BibitemShut {NoStop}%
\bibitem [{\citenamefont {Floreanini}\ and\ \citenamefont
  {Jackiw}(1987)}]{Jackiw}%
  \BibitemOpen
  \bibfield  {author} {\bibinfo {author} {\bibfnamefont {R.}~\bibnamefont
  {Floreanini}}\ and\ \bibinfo {author} {\bibfnamefont {R.}~\bibnamefont
  {Jackiw}},\ }\href {\doibase 10.1103/PhysRevLett.59.1873} {\bibfield
  {journal} {\bibinfo  {journal} {Phys. Rev. Lett.}\ }\textbf {\bibinfo
  {volume} {59}},\ \bibinfo {pages} {1873} (\bibinfo {year}
  {1987})}\BibitemShut {NoStop}%
\bibitem [{\citenamefont {Coussaert}\ \emph {et~al.}(1995)\citenamefont
  {Coussaert}, \citenamefont {Henneaux},\ and\ \citenamefont {van
  Driel}}]{Coussaert:1995zp}%
  \BibitemOpen
  \bibfield  {author} {\bibinfo {author} {\bibfnamefont {O.}~\bibnamefont
  {Coussaert}}, \bibinfo {author} {\bibfnamefont {M.}~\bibnamefont {Henneaux}},
  \ and\ \bibinfo {author} {\bibfnamefont {P.}~\bibnamefont {van Driel}},\
  }\href {\doibase 10.1088/0264-9381/12/12/012} {\bibfield  {journal} {\bibinfo
   {journal} {Class. Quant. Grav.}\ }\textbf {\bibinfo {volume} {12}},\
  \bibinfo {pages} {2961} (\bibinfo {year} {1995})},\ \Eprint
  {http://arxiv.org/abs/gr-qc/9506019} {arXiv:gr-qc/9506019} \BibitemShut
  {NoStop}%
\bibitem [{\citenamefont {Cartan}(1923)}]{Cartan:1923zea}%
  \BibitemOpen
  \bibfield  {author} {\bibinfo {author} {\bibfnamefont {E.}~\bibnamefont
  {Cartan}},\ }\href@noop {} {\bibfield  {journal} {\bibinfo  {journal}
  {Annales Sci. Ecole Norm. Sup.}\ }\textbf {\bibinfo {volume} {40}},\ \bibinfo
  {pages} {325} (\bibinfo {year} {1923})}\BibitemShut {NoStop}%
\bibitem [{\citenamefont {L{\'e}vy-Leblond}(1971)}]{levy1971galilei}%
  \BibitemOpen
  \bibfield  {author} {\bibinfo {author} {\bibfnamefont {J.-M.}\ \bibnamefont
  {L{\'e}vy-Leblond}},\ }in\ \href@noop {} {\emph {\bibinfo {booktitle} {Group
  theory and its applications}}}\ (\bibinfo  {publisher} {Elsevier},\ \bibinfo
  {year} {1971})\ pp.\ \bibinfo {pages} {221--299}\BibitemShut {NoStop}%
\bibitem [{\citenamefont {Grigore}(1996)}]{Grigore:1993fz}%
  \BibitemOpen
  \bibfield  {author} {\bibinfo {author} {\bibfnamefont {D.~R.}\ \bibnamefont
  {Grigore}},\ }\href {\doibase 10.1063/1.531402} {\bibfield  {journal}
  {\bibinfo  {journal} {J. Math. Phys.}\ }\textbf {\bibinfo {volume} {37}},\
  \bibinfo {pages} {460} (\bibinfo {year} {1996})},\ \Eprint
  {http://arxiv.org/abs/hep-th/9312048} {arXiv:hep-th/9312048} \BibitemShut
  {NoStop}%
\bibitem [{\citenamefont {Bose}(1995)}]{Bose:1994sj}%
  \BibitemOpen
  \bibfield  {author} {\bibinfo {author} {\bibfnamefont {S.~K.}\ \bibnamefont
  {Bose}},\ }\href {\doibase 10.1007/BF02099478} {\bibfield  {journal}
  {\bibinfo  {journal} {Commun. Math. Phys.}\ }\textbf {\bibinfo {volume}
  {169}},\ \bibinfo {pages} {385} (\bibinfo {year} {1995})}\BibitemShut
  {NoStop}%
\bibitem [{\citenamefont {Bergshoeff}\ and\ \citenamefont
  {Rosseel}(2016)}]{Bergshoeff:2016lwr}%
  \BibitemOpen
  \bibfield  {author} {\bibinfo {author} {\bibfnamefont {E.~A.}\ \bibnamefont
  {Bergshoeff}}\ and\ \bibinfo {author} {\bibfnamefont {J.}~\bibnamefont
  {Rosseel}},\ }\href {\doibase 10.1103/PhysRevLett.116.251601} {\bibfield
  {journal} {\bibinfo  {journal} {Phys. Rev. Lett.}\ }\textbf {\bibinfo
  {volume} {116}},\ \bibinfo {pages} {251601} (\bibinfo {year} {2016})},\
  \Eprint {http://arxiv.org/abs/1604.08042} {arXiv:1604.08042 [hep-th]}
  \BibitemShut {NoStop}%
\bibitem [{\citenamefont {Avil\'es}\ \emph {et~al.}(2018)\citenamefont
  {Avil\'es}, \citenamefont {Frodden}, \citenamefont {Gomis}, \citenamefont
  {Hidalgo},\ and\ \citenamefont {Zanelli}}]{Aviles:2018jzw}%
  \BibitemOpen
  \bibfield  {author} {\bibinfo {author} {\bibfnamefont {L.}~\bibnamefont
  {Avil\'es}}, \bibinfo {author} {\bibfnamefont {E.}~\bibnamefont {Frodden}},
  \bibinfo {author} {\bibfnamefont {J.}~\bibnamefont {Gomis}}, \bibinfo
  {author} {\bibfnamefont {D.}~\bibnamefont {Hidalgo}}, \ and\ \bibinfo
  {author} {\bibfnamefont {J.}~\bibnamefont {Zanelli}},\ }\href {\doibase
  10.1007/JHEP05(2018)047} {\bibfield  {journal} {\bibinfo  {journal} {JHEP}\
  }\textbf {\bibinfo {volume} {05}},\ \bibinfo {pages} {047} (\bibinfo {year}
  {2018})},\ \Eprint {http://arxiv.org/abs/1802.08453} {arXiv:1802.08453
  [hep-th]} \BibitemShut {NoStop}%
\bibitem [{\citenamefont {Gonz\'alez}\ \emph {et~al.}(2016)\citenamefont
  {Gonz\'alez}, \citenamefont {Rubio}, \citenamefont {Salgado},\ and\
  \citenamefont {Salgado}}]{Gonzalez:2016xwo}%
  \BibitemOpen
  \bibfield  {author} {\bibinfo {author} {\bibfnamefont {N.}~\bibnamefont
  {Gonz\'alez}}, \bibinfo {author} {\bibfnamefont {G.}~\bibnamefont {Rubio}},
  \bibinfo {author} {\bibfnamefont {P.}~\bibnamefont {Salgado}}, \ and\
  \bibinfo {author} {\bibfnamefont {S.}~\bibnamefont {Salgado}},\ }\href
  {\doibase 10.1016/j.physletb.2016.02.037} {\bibfield  {journal} {\bibinfo
  {journal} {Phys. Lett. B}\ }\textbf {\bibinfo {volume} {755}},\ \bibinfo
  {pages} {433} (\bibinfo {year} {2016})},\ \Eprint
  {http://arxiv.org/abs/1604.06313} {arXiv:1604.06313 [hep-th]} \BibitemShut
  {NoStop}%
\bibitem [{\citenamefont {Gomis}\ \emph {et~al.}(2019)\citenamefont {Gomis},
  \citenamefont {Kleinschmidt},\ and\ \citenamefont
  {Palmkvist}}]{Gomis:2019fdh}%
  \BibitemOpen
  \bibfield  {author} {\bibinfo {author} {\bibfnamefont {J.}~\bibnamefont
  {Gomis}}, \bibinfo {author} {\bibfnamefont {A.}~\bibnamefont {Kleinschmidt}},
  \ and\ \bibinfo {author} {\bibfnamefont {J.}~\bibnamefont {Palmkvist}},\
  }\href {\doibase 10.1007/JHEP09(2019)109} {\bibfield  {journal} {\bibinfo
  {journal} {JHEP}\ }\textbf {\bibinfo {volume} {09}},\ \bibinfo {pages} {109}
  (\bibinfo {year} {2019})},\ \Eprint {http://arxiv.org/abs/1907.00410}
  {arXiv:1907.00410 [hep-th]} \BibitemShut {NoStop}%
\bibitem [{\citenamefont {Concha}\ and\ \citenamefont
  {Rodr\'\i{}guez}(2019)}]{Concha:2019lhn}%
  \BibitemOpen
  \bibfield  {author} {\bibinfo {author} {\bibfnamefont {P.}~\bibnamefont
  {Concha}}\ and\ \bibinfo {author} {\bibfnamefont {E.}~\bibnamefont
  {Rodr\'\i{}guez}},\ }\href {\doibase 10.1007/JHEP07(2019)085} {\bibfield
  {journal} {\bibinfo  {journal} {JHEP}\ }\textbf {\bibinfo {volume} {07}},\
  \bibinfo {pages} {085} (\bibinfo {year} {2019})},\ \Eprint
  {http://arxiv.org/abs/1906.00086} {arXiv:1906.00086 [hep-th]} \BibitemShut
  {NoStop}%
\bibitem [{\citenamefont {Soroka}\ and\ \citenamefont
  {Soroka}(2009)}]{Soroka:2006aj}%
  \BibitemOpen
  \bibfield  {author} {\bibinfo {author} {\bibfnamefont {D.~V.}\ \bibnamefont
  {Soroka}}\ and\ \bibinfo {author} {\bibfnamefont {V.~A.}\ \bibnamefont
  {Soroka}},\ }\href {\doibase 10.1155/2009/234147} {\bibfield  {journal}
  {\bibinfo  {journal} {Adv. High Energy Phys.}\ }\textbf {\bibinfo {volume}
  {2009}},\ \bibinfo {pages} {234147} (\bibinfo {year} {2009})},\ \Eprint
  {http://arxiv.org/abs/hep-th/0605251} {arXiv:hep-th/0605251} \BibitemShut
  {NoStop}%
\bibitem [{\citenamefont {Gibbons}\ \emph {et~al.}(2010)\citenamefont
  {Gibbons}, \citenamefont {Gomis},\ and\ \citenamefont
  {Pope}}]{Gibbons:2009me}%
  \BibitemOpen
  \bibfield  {author} {\bibinfo {author} {\bibfnamefont {G.~W.}\ \bibnamefont
  {Gibbons}}, \bibinfo {author} {\bibfnamefont {J.}~\bibnamefont {Gomis}}, \
  and\ \bibinfo {author} {\bibfnamefont {C.~N.}\ \bibnamefont {Pope}},\ }\href
  {\doibase 10.1103/PhysRevD.82.065002} {\bibfield  {journal} {\bibinfo
  {journal} {Phys. Rev. D}\ }\textbf {\bibinfo {volume} {82}},\ \bibinfo
  {pages} {065002} (\bibinfo {year} {2010})},\ \Eprint
  {http://arxiv.org/abs/0910.3220} {arXiv:0910.3220 [hep-th]} \BibitemShut
  {NoStop}%
\bibitem [{\citenamefont {Salgado-Rebolledo}()}]{Salgado-Rebolledo:2019kft}%
  \BibitemOpen
  \bibfield  {author} {\bibinfo {author} {\bibfnamefont {P.}~\bibnamefont
  {Salgado-Rebolledo}},\ }\href {\doibase 10.1007/JHEP10(2019)039} {\bibfield
  {journal} {\bibinfo  {journal} {JHEP}\ }\textbf {\bibinfo {volume} {10}},\
  \bibinfo {pages} {039}},\ \Eprint {http://arxiv.org/abs/1905.09421}
  {arXiv:1905.09421 [hep-th]} \BibitemShut {NoStop}%
\bibitem [{\citenamefont {Cappelli}\ and\ \citenamefont
  {Maffi}(2018)}]{Cappelli3}%
  \BibitemOpen
  \bibfield  {author} {\bibinfo {author} {\bibfnamefont {A.}~\bibnamefont
  {Cappelli}}\ and\ \bibinfo {author} {\bibfnamefont {L.}~\bibnamefont
  {Maffi}},\ }\href {\doibase 10.1088/1751-8121/aad0ab} {\bibfield  {journal}
  {\bibinfo  {journal} {Journal of Physics A: Mathematical and Theoretical}\
  }\textbf {\bibinfo {volume} {51}},\ \bibinfo {pages} {365401} (\bibinfo
  {year} {2018})}\BibitemShut {NoStop}%
\bibitem [{\citenamefont {Golkar}\ \emph
  {et~al.}(2016{\natexlab{b}})\citenamefont {Golkar}, \citenamefont {Nguyen},
  \citenamefont {Roberts},\ and\ \citenamefont {Son}}]{Son4}%
  \BibitemOpen
  \bibfield  {author} {\bibinfo {author} {\bibfnamefont {S.}~\bibnamefont
  {Golkar}}, \bibinfo {author} {\bibfnamefont {D.~X.}\ \bibnamefont {Nguyen}},
  \bibinfo {author} {\bibfnamefont {M.~M.}\ \bibnamefont {Roberts}}, \ and\
  \bibinfo {author} {\bibfnamefont {D.~T.}\ \bibnamefont {Son}},\ }\href
  {\doibase 10.1103/PhysRevLett.117.216403} {\bibfield  {journal} {\bibinfo
  {journal} {Phys. Rev. Lett.}\ }\textbf {\bibinfo {volume} {117}},\ \bibinfo
  {pages} {216403} (\bibinfo {year} {2016}{\natexlab{b}})}\BibitemShut
  {NoStop}%
\bibitem [{\citenamefont {Liu}\ \emph {et~al.}(2018)\citenamefont {Liu},
  \citenamefont {Gromov},\ and\ \citenamefont {Papi\ifmmode~\acute{c}\else
  \'{c}\fi{}}}]{Papic}%
  \BibitemOpen
  \bibfield  {author} {\bibinfo {author} {\bibfnamefont {Z.}~\bibnamefont
  {Liu}}, \bibinfo {author} {\bibfnamefont {A.}~\bibnamefont {Gromov}}, \ and\
  \bibinfo {author} {\bibfnamefont {Z.}~\bibnamefont
  {Papi\ifmmode~\acute{c}\else \'{c}\fi{}}},\ }\href {\doibase
  10.1103/PhysRevB.98.155140} {\bibfield  {journal} {\bibinfo  {journal} {Phys.
  Rev. B}\ }\textbf {\bibinfo {volume} {98}},\ \bibinfo {pages} {155140}
  (\bibinfo {year} {2018})}\BibitemShut {NoStop}%
\end{thebibliography}%






\end{document}